\documentclass[conference]{IEEEtran}
\IEEEoverridecommandlockouts
\usepackage{cite}
\usepackage{amsmath,amssymb,amsfonts}
\usepackage{algorithmic}
\usepackage{graphicx}
\usepackage{textcomp}
\usepackage{xcolor}

\usepackage{array}
\usepackage{booktabs}
\usepackage{multirow}

\usepackage{pifont}
\usepackage{enumitem}
\setlist[itemize]{leftmargin=*}
\setlist[enumerate]{leftmargin=*}

\usepackage[skip=2pt,font=footnotesize,labelformat=simple]{caption,subcaption}
\usepackage{balance}

\makeatletter
\newcommand{\linebreakand}{%
  \end{@IEEEauthorhalign}
  \hfill\mbox{}\par
  \mbox{}\hfill\begin{@IEEEauthorhalign}
}
\makeatother

\usepackage[misc]{ifsym}
\usepackage{hyperref}

\def\BibTeX{{\rm B\kern-.05em{\sc i\kern-.025em b}\kern-.08em
    T\kern-.1667em\lower.7ex\hbox{E}\kern-.125emX}}
\begin{document}

\newcommand{\myFrame}{SimDC}
\title{SimDC: A High-Fidelity Device Simulation Platform for Device-Cloud Collaborative Computing}


\let\SUP\textsuperscript
\author{
\IEEEauthorblockN{Ruiguang Pei\IEEEauthorrefmark{1}\thanks{\IEEEauthorrefmark{1} Both authors contributed equally to this research.}}
\IEEEauthorblockA{OPPO Research Institute\\
peiruiguang@gmail.com}
\and
\IEEEauthorblockN{Junjie Wu\IEEEauthorrefmark{1}}
\IEEEauthorblockA{OPPO Research Institute\\
wujunjiepku@163.com}
\and
\IEEEauthorblockN{Dan Peng}
\IEEEauthorblockA{OPPO Research Institute\\
lepengdan@gmail.com}
\linebreakand 
\IEEEauthorblockN{Min Fang}
\IEEEauthorblockA{OPPO Research Institute\\
mfang.cs@gmail.com}
\and
\IEEEauthorblockN{Jianan Zhang}
\IEEEauthorblockA{OPPO Research Institute\\
zmanda\_zhang@163.com}
\and
\IEEEauthorblockN{Zhihui Fu\IEEEauthorrefmark{2}}
\IEEEauthorblockA{OPPO Research Institute\\
luca@oppo.com}
\and
\IEEEauthorblockN{Jun Wang\IEEEauthorrefmark{2}\thanks{\IEEEauthorrefmark{2} Corresponding author.}}
\IEEEauthorblockA{OPPO Research Institute\\
junwang.lu@gmail.com}
}
\maketitle

\begin{abstract}
The advent of edge intelligence and escalating concerns for data privacy protection have sparked a surge of interest in device-cloud collaborative computing. 
Large-scale device deployments to validate prototype solutions are often prohibitively expensive and practically challenging, resulting in a pronounced demand for simulation tools that can emulate real-world scenarios.
However, existing simulators predominantly rely solely on high-performance servers to emulate edge computing devices, overlooking (1) the discrepancies between virtual computing units and actual heterogeneous computing devices and (2) the simulation of device behaviors in real-world environments.
In this paper, we propose a high-fidelity device simulation platform, called \textbf{\myFrame}, which uses a hybrid heterogeneous resource and integrates high-performance servers and physical mobile phones. 
Utilizing this platform, developers can simulate numerous devices for functional testing cost-effectively and capture precise operational responses from varied real devices.
To simulate real behaviors of heterogeneous devices, we offer a configurable device behavior traffic controller that dispatches results on devices to the cloud using a user-defined operation strategy.
Comprehensive experiments on the public dataset show the effectiveness of our simulation platform and its great potential for application.
The code is available at \url{https://github.com/opas-lab/olearning-sim}.
\end{abstract}


\section{Introduction}
{\color{black}
Cross-device collaborative computing, such as federated learning (FL)~\cite{fl}, has emerged as a promising solution to meet the increasing demand for low-latency, high-throughput, and privacy-preserving applications~\cite{papaya,fate,federatedscope,appleODL}. 
However, due to the large-scale deployment, heterogeneity of edge devices, and the diversity of their operational behaviors, the development and evaluation of device-cloud collaborative computing algorithms and applications face significant challenges~\cite{fedjax,flute,flsim}. 
Therefore, it is crucial to design and implement a high-fidelity device simulation platform for device-cloud collaborative computing, which provides a controlled and reproducible environment for real-world deployment. With this simulator, developers can quickly understand the operational effectiveness of their collaborative computing algorithms and applications under various conditions at a relatively low cost.

Most current simulators~\cite{privacyfl,fedjax,flute,flsim,fedml,flower,fsrealvldb,fsrealkdd,roth2022nvidia,zhou2022role} predominantly rely on high-performance servers. These systems frequently utilize multi-threading, multi-processing, and distributed computing techniques, creating separate processes to represent devices within device-cloud collaborative computing frameworks. This methodology is highly beneficial for addressing substantial challenges in large-scale device-cloud collaborative simulation scenarios. 
{\color{black}It allows extending the simulation scope to tens of thousands of devices without significant difficulties.}
Assuming sufficient computational resources are accessible, this strategy enables the efficient simulation and validation of 
{\color{black} device-cloud collaborative computing algorithms,} providing essential experimental data and insights for analysis.

However, there is still a gap between such simulation methods and real device-cloud {\color{black}collaborative computing} scenarios. The challenges encountered in real-world contexts can be categorized into two main areas:
\begin{enumerate}
    \item \textbf{The discrepancy between virtual computing devices and actual edge devices.} Although the scalability of virtual computing resources (based on high-performance servers) can meet the needs of numerous edge devices, they do not accurately emulate the real-time responses or operational burdens of physical devices, such as energy efficiency. Furthermore, the heterogeneity of edge devices (e.g., smartphones ranging from high to low configurations) introduces differences in communication efficacy and computational latency. These disparities are pivotal in implementing {\color{black}device-cloud collaborative computing} applications on edge devices. For example, a device is permitted to participate in collaborative computing only if the user experience remains unaffected, this will ultimately impact the overall system and applications~\cite{hard2018federated,yang2019federated}.
    \item {\color{black}\textbf{The divergence between edge device behavior in simulations and the real world.}} The behavior of edge devices within device-cloud {\color{black}collaborative computing} simulations is often underestimated, but in practical deployments, they significantly diverge due to variables like time zones and network connectivity. These discrepancies can considerably influence both the efficiency of collaborative computing and the stability of the system~\cite{bonawitz2019towards,luo2021cost}. The ramifications of such impacts can be dissected into system-level and application-level concerns. For the former, the cloud infrastructure (e.g., aggregation services) may experience fluctuating access loads due to the uneven distribution of devices across different localities. For the latter, network inconsistencies and disconnections driven by diverse regional factors can compromise the algorithm's ultimate accuracy.
\end{enumerate}

Recent advances~\cite{fsrealkdd,fsrealvldb,woisetschlager2024fledge} have highlighted the importance of accurately reflecting real-world scenarios in device-cloud collaborative computing simulations. 
Nevertheless, a truly holistic system that concurrently satisfies the demands of extensive simulations, authentic device emulation, device diversity, and accurate representation of edge device behaviors remains absent. Such a system would closely replicate real-world device-cloud collaborative computing settings in a comprehensive end-to-end simulation.

In light of these considerations, we introduce a device-cloud collaborative {\color{black}computing} simulation system tailored for real-world scenarios {\color{black}(denoted as \myFrame)}, grounded on the following foundational principles: (1) \textbf{Scalability}: Acknowledging that real-world conditions frequently entail handling requests from millions, even tens of millions, of devices. (2) \textbf{Heterogeneous Computing Devices}: Our system integrates actual edge devices while also ensuring a diverse spectrum of device capabilities is represented. (3) \textbf{Simulation of Edge Device Behavior}: We incorporate the realistic behavior of edge devices, acknowledging its substantial influence on the overall performance of the application.

To this end, we have implemented a hybrid heterogeneous resource framework, comprising: \emph{High-Performance Server Cluster}, which is employed to cost-effectively simulate an extensive array of devices for functional validation, paralleling approaches from existing literature; and \emph{Physical Mobile Phone Cluster}, consisting of Android mobile phones with varied configurations to capture authentic operational responses during algorithm training.
In conjunction with this, we have developed an adept device resource coordination mechanism, tailored for our hybrid resource clusters, aimed at minimizing temporal costs while catering to the heterogeneous simulation demands of users.
Further ensuring the fidelity of device behavior simulation, our system includes a bespoke device behavior traffic controller named \emph{DeviceFlow}.
As previously analyzed, the behaviors of devices primarily exert system pressure on cloud services or influence the overall outcomes. From the perspective of cloud services, device behaviors are mainly characterized by fluctuations in request traffic and disconnections. Based on this, DeviceFlow simulates device behaviors primarily by controlling the variations in network traffic from the edge to the cloud, in line with user-defined operational parameters.
The major contributions of our paper are summarized as follows:
\begin{itemize}
    \item We propose a towards real-world scenarios device simulation platform for device-cloud collaboration computing, which employs a hybrid heterogeneous cluster design to support low-cost evaluation of algorithmic correctness and obtain real physical operating responses. Meanwhile, we design an efficient device allocation strategy to maximize resource utilization and minimize time costs. 
    \item We implement a novel programmable device behavior traffic controller, which provides a simulation of the varied behavior of heterogeneous edge devices in real-world environments according to user-specified operation specifications.  
    \item Extensive experiments indicate excellent scalability (easily scalable to support a large number of simulation devices), as well as high accuracy and efficiency of the designed simulation platform.
\end{itemize}

}

\section{Related Work}
\label{sec:relatedWork}
{\color{black}
\subsection{Device-cloud Collaborative Computing}
\textcolor{black}{Federated Learning (FL), as a typical device-cloud collaborative computing paradigm,}
has garnered considerable interest since its emergence due to its ability to compute collaboratively while keeping each device's raw data stored locally~\cite{papaya,fate,federatedscope}. FL operates as a distributed optimization challenge, where numerous clients, such as smartphones and IoT devices, collectively refine a global model. Here, each client harbors private data on-device and contributes to model training by retaining data proximity. The primary objective of FL is to determine model parameters that minimize the overall loss function across all devices. The process unfolds as follows:
(1) \emph{Client Set}: Define $\mathcal{K}$ to be the set of clients engaging in federated learning, where each client $k \in \mathcal{K}$ maintains a local dataset $\mathcal{D}_k$.
(2) \emph{Global Model Parameters}: Denote by $\mathbf{w}$ the parameters for the global model.
(3) \emph{Local Loss Function}: For each client $k$, ascertain the local loss function, $F_k(\mathbf{w}; \mathcal{D}_k)$, to assess the performance of the model using parameters $\mathbf{w}$ on the local dataset $\mathcal{D}_k$.
(4) \emph{Optimization Objective}: The federated learning endeavor seeks to minimize the global loss function:
$$\min_{\mathbf{w}} F(\mathbf{w}) = \sum_{k \in \mathcal{K}} p_k F_k(\mathbf{w}; \mathcal{D}k)$$
where $p_k$ signifies the weight assigned to client $k$, under the constraint $\sum_{k \in \mathcal{K}} p_k = 1$. Typically, $p_k$ is proportionate to the size of the dataset held by client $k$.
Throughout the training regimen, local devices relay their individually trained models to the cloud infrastructure for integration, utilizing aggregation strategies like FedAvg~\cite{fl}. The merged global model is then redistributed to the local devices to facilitate successive rounds of local refinement. This cyclical interaction between local computation and global aggregation persists until the model attains convergence at round $T$.

For purposes like communication cost or personalization, the efficiency-primary co-computing paradigm is proposed, where the cloud can collect non-privacy concerns historical features~\cite{sigiredagecloud,yao2022device,yao2021device,edgerec}.
In practical applications, the data distribution across different devices may exhibit significant variations(non-IID), which have been explored in the context of optimizations for personalized federated learning~\cite{li2021ditto,li2021fedbn,paulik2021federated,shin2022fedbalancer}.
Partial training-based~\cite{fjord,fedrolex,autosplit,neurosurgeon} and knowledge distillation-based~\cite{fedet,fedgkt,fedagg} methods enhance co-computing capabilities between edge devices and cloud.

\subsection{Device-cloud Collaborative Computing Simulators}
Simulators, such as FedML~\cite{fedml} and Flower~\cite{flower}, serve mainly for proof-of-concept in the production process of specific frameworks.
\cite{privacyfl} focuses on configurable privacy scenarios and device behavior simulation under static networks.
Google's FEDJAX~\cite{fedjax}, Microsoft's FLUTE~\cite{flute}, Meta's FLSim~\cite{flsim}, NVIDIA FLARE~\cite{roth2022nvidia}, Fedscale~\cite{lai2022fedscale}, FedratedScope~\cite{federatedscope}, and Pysyft~\cite{ziller2021pysyft} are standalone simulations under high-performance servers, considering only homogeneous devices.
At the hardware level, these simulation frameworks harness high-performance servers equipped with CPUs or GPUs to emulate computation, making use of distributed computing's scalability to represent a broad array of edge computing devices. This approach facilitates large-scale simulations of device-cloud collaborative computing. Yet, these frameworks exhibit certain limitations:
(1) \textbf{Simulation of Heterogeneous Computing Devices}: In actual use cases, there is a natural heterogeneity and diversity among devices (from low to high-end smartphones), leading to variances in their computational and communication capacities that are crucial for appraising the performance of individual devices.
(2) \textbf{Simulation of Edge Device Behavior}: Overlooking the behaviors of edge computing devices, such as network conditions, disconnections, and application crashes, can detrimentally affect both local and global performance assessments.

To tackle these challenges, some works like FS-Real\cite{fsrealvldb,fsrealkdd} embarked on an assessment of current device-cloud collaborative algorithms across various heterogeneous hardware platforms. This research revealed significant disparities between the performance of heterogeneous devices and the often assumed homogeneity in real settings. As a result, authors introduced a logic-based simulation technique for FL that accounts for the heterogeneity of edge devices.
Further, FLEdge\cite{woisetschlager2024fledge} enhances existing FL benchmarks by systematically evaluating client competencies, with particular emphasis on client behaviors and the impact of computational and communication bottlenecks. 
Though these initiatives mark a foray into exploring the previously noted limitations, FS-Real still approaches the heterogeneity of edge devices using Android native simulation tools (which operate on HPCs), and does not capture the real-time reactions of actual devices. Meanwhile, FLEdge focuses chiefly on benchmarking assessments.

Presently, a comprehensive and user-friendly \textcolor{black}{device simulation}
that can holistically address these issues in an end-to-end fashion is yet to be developed.
}
\section{Architecture Overview}\label{sec:architecture_overview}

The SimDC platform is designed to provide a high-fidelity simulation platform for simulating the calculation and behavior of edge devices within device-cloud collaboration scenarios. This platform facilitates the advancement of cloud-side services and edge-side algorithms by closely replicating real-world application contexts. Compared to other similar simulators,
SimDC offers two corresponding advantages:

\textbf{Leveraging Heterogeneous Hybrid Resources.} This platform integrates high-performance computing resources to support large-scale simulations of edge devices, while also incorporating a physical devices cluster. This configuration accurately captures critical physical performance metrics, including power consumption, CPU and memory usage, for edge devices executing specific algorithms during task execution. 
This capability provides substantial support for selecting and managing edge devices based on actual application scenarios.

\textbf{Controlling Large-Scale Device Behavior Traffic.} By constructing a programmable device behavior traffic controller, the SimDC platform is capable of simulating large-scale behavior patterns of edge groups during device-cloud collaboration, providing highly realistic information and data streams for the development and testing of cloud-side services.

\subsection{Task Design Specifications}

In SimDC platform, a task serves as the core operational unit, adhering to strict design specifications to ensure its validity and accuracy. Each task is assigned a unique identifier (task\_id), allowing for precise identification and tracking to enhance efficiency in task management and monitoring. Users can simulate multiple devices with varying performance levels within a single task, all of which must execute the same computational process (operator flow) uniformly.

The task design is based on a singular operator flow, composed of multiple operators in a predetermined sequence. A task allows simulated devices to repetitively execute the same operator flow multiple times, thereby enabling the simulation of multi-round device-cloud collaborative processes. Within a single operator flow, various datasets can be employed. It is necessary to explicitly specify the corresponding device grades and the total number of devices to be simulated for different datasets, which ensures the accurate reflection of the performance characteristics of simulation devices across different performance levels and scales during the simulation.

Additionally, a task can request heterogeneous hybrid resources for computation, enhancing the adaptability and complexity of task execution. Each task can also be configured with a "scheduling priority" parameter that determines the order in which tasks are executed if resources are sufficient. 
\begin{figure}[t]
  \centering
  \includegraphics[width=\linewidth]{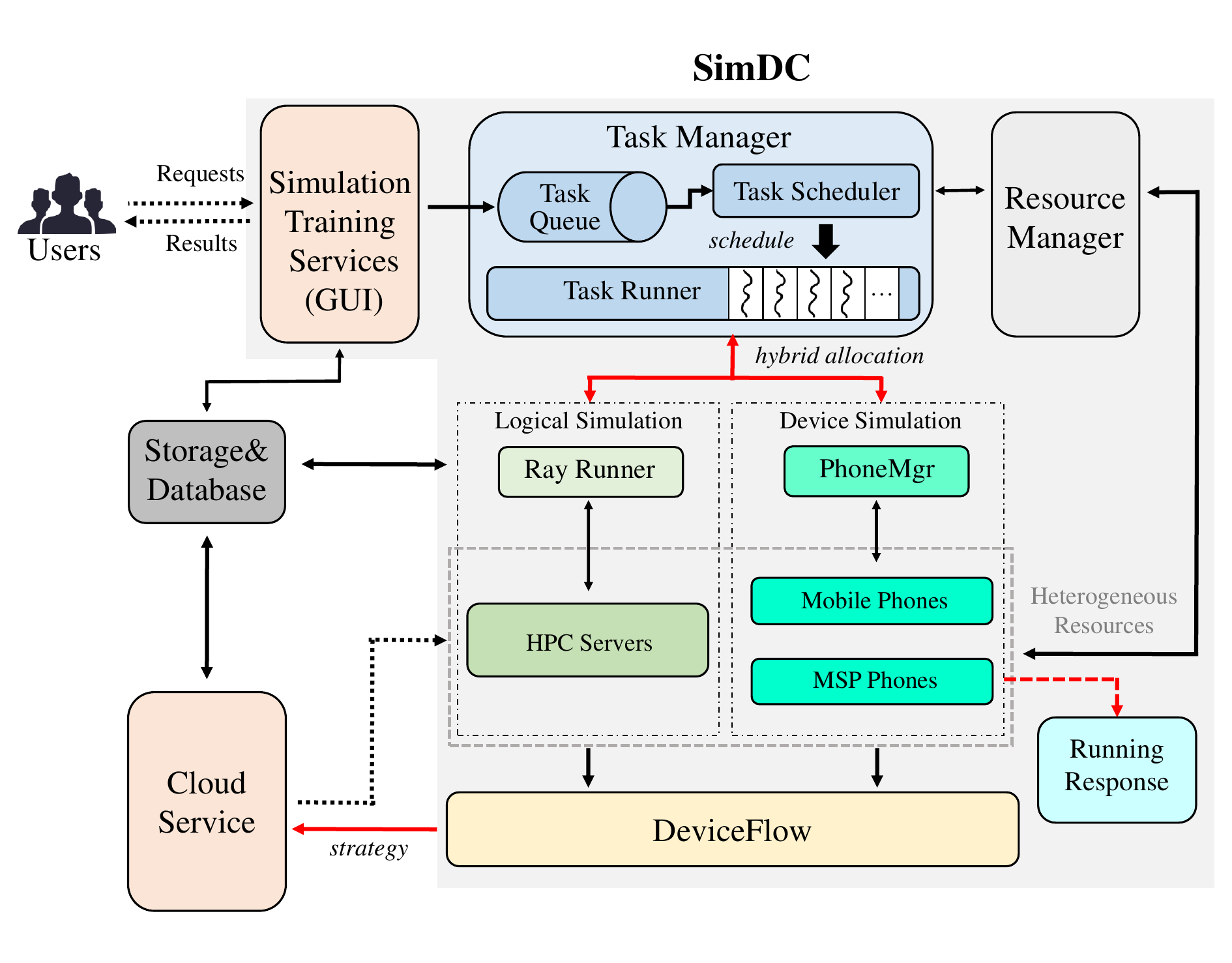}
  \caption{Architectural schematic of the SimDC platform.}
  \label{fig:simdc_overview}
\end{figure}

\subsection{Platform Architecture}

SimDC contains several key modules that work together to implement its core functionality, as shown in Fig.~\ref{fig:simdc_overview}, including:

\begin{itemize}
    \item \emph{Task Manager}. A micro-service responsible for the maintenance of \emph{Task Queue}, task submission, and status monitoring. \emph{Task Manager} periodically selects suitable submitted tasks from the \emph{Task Queue} for scheduling, leveraging two main sub-components: \emph{Task Scheduler} and \emph{Task Runner}. \emph{Task Scheduler} employs a greedy algorithm to schedule tasks from the queue, taking into account the current states of the resource pool from \emph{Resource Manager}, demand resources, and the expected task benefits derived from the scheduling priority. It prioritizes tasks that meet resource requirements while maximizing the anticipated benefits. \emph{Task Runner} dynamically adjusts execution strategies for scheduled tasks, ensuring that they are allocated to appropriate heterogeneous resources based on the requested resource amounts and the number of simulated devices. Additionally, the \emph{Task Runner} supports multi-threaded concurrent processing to optimize task execution efficiency.
    \item \emph{Resource Manager}. This module oversees the querying, freezing, and releasing of heterogeneous resources, while also enabling dynamic scaling up or down. \emph{Resource Manager} continuously monitors physical resources in real-time and synchronizes resource utilization information with the \emph{Task Manager} to facilitate efficient task control.
    \item \emph{Logical Simulation}. By utilizing high-performance server resources, \emph{Logical Simulation} emulates the computational processes of edge devices, offering high flexibility and scalability. This module can dynamically adjust the allocation of computing resources according to task requirements and supports user-defined operators, thus being able to adapt to various types and complexity of simulation tasks.
    \item \emph{Device Simulation}. \emph{Device Simulation} includes the phone devices management module (\emph{PhoneMgr}) and a physical devices cluster. \emph{PhoneMgr} is responsible for selecting appropriate real phone devices to participate in the simulation based on task requirements. It manages task submission, status monitoring, termination operations, and performance measurement. The physical devices cluster can be configured and scaled up or down according to demand.
    \item \emph{DeviceFlow}. \emph{DeviceFlow} serves as a device behavior traffic controller. It buffers intermediate information or related data generated by simulated edge computations and sends them to the cloud based on user-defined traffic strategies. This functionality emulates various traffic scenarios of large-scale edge responses directed at \emph{Cloud Service}, accurately reflecting the collective behavior of devices.
\end{itemize}

The collaborative operation of these components ensures that every stage of task management, from creation to execution, is conducted efficiently and accurately. This provides users with a robust and flexible device simulation platform.
\subsection{Task Execution Workflow}

Once users set device simulation targets, cloud service parameters, resource requirements, and operator flow configurations via the front-end graphical user interface (GUI), they can create and submit tasks. These tasks are added to the \emph{Task Queue} for processing. The \emph{Task Manager} then invokes the \emph{Resource Manager} to schedule tasks based on the current availability of resources and task priorities, optimizing the allocation of requested heterogeneous resources according to user requirements. Subsequently, the tasks are distributed to the \emph{Logical Simulation} and \emph{Device Simulation} for execution.

If the user's task requires simulating traffic scenarios for edge device behavior, the interaction-related information or data with the cloud side are temporarily stored in \emph{DeviceFlow}. This data is then transmitted to the \emph{Cloud Service} according to the user-defined traffic strategy. Ultimately, users can monitor various computational metrics, edge device performance, and updates to cloud services throughout the task execution process via the GUI, until the task is completed.
\section{Hybrid Heterogeneous Resources}

\subsection{Heterogeneous Resources}

The heterogeneous resources framework is designed to provide an efficient, flexible, and easily manageable simulation environment by integrating simulated devices in \emph{Logical Simulation} and physical devices in \emph{Device Simulation}. For \emph{Logical Simulation}, the platform leverages high-performance servers and employs Kubernetes (k8s) nodes for elastic scaling to accommodate simulation demands of varying scales. Specifically, Ray clusters are deployed on k8s nodes, and the platform utilizes the Ray's job submission mechanism to execute specific computational tasks with operator flow. The master node (\emph{Ray Runner}) is responsible for data downloading, distribution, and the configuration of runtime parameters for the simulated devices. Subsequently, this master node utilizes Ray's distributed computing framework to directly launch placement groups of actors on worker nodes, with each actor sequentially simulating multiple devices. This approach enables efficient large-scale simulation for devices.

\emph{Device Simulation} developed \emph{PhoneMgr} module to manage tasks on physical phone devices. \emph{PhoneMgr} first handles the downloading and distribution of data, then employs Android Debug Bridge (ADB) commands to directly control the execution process of phone devices in the physical devices cluster. The physical devices cluster includes local mobile phones and remote phones provided by the Mobile Service Platform (MSP). These devices vary in model and configuration, allowing users to categorize them based on customizable settings, such as performance levels (e.g., High, Low) or specific device models. Notably, the physical devices cluster can designate certain phones for performance measurement during task execution, enabling the collection of runtime parameters to facilitate the development and tuning of edge algorithms.


\subsection{Hybrid Allocation Optimization}

SimDC allows tasks to be configured with hybrid heterogeneous resources for simulating devices. A task in \emph{Task Runner} scheduled by \emph{Task Scheduler} comprises user's simulation requirements. Based on the total number of simulation devices and computing resources, \emph{Task Runner} determines the optimal allocation of device quantities across hybrid heterogeneous resources, aiming to minimize the execution time. 

Assuming that the user needs to simulate $c$ grades of devices with quantities $\{N_1, N_2, \ldots, N_c\}$, of which $\{q_1, q_2, \ldots, q_c\}$ are physical phone devices reserved for measuring performance metrics. The required resources for each grade of devices include $\{f_1, f_2, \ldots, f_c\}$ bundles in \emph{Logical Simulation}, as well as $\{m_1, m_2, \ldots, m_c\}$ physical phone devices in \emph{Device Simulation}. In \emph{Logical Simulation}, $\{k_1, k_2, \ldots, k_c\}$ represent the unit resource bundles, which include CPU, memory or GPU, used for emulating devices. For instance, a unit resource bundle is defined as $\{$CPU: 1 core, memory: 1 GB$\}$, and a High-grade device requires 8 such units to simulate ($k=8$), while the user requires 10 High-grade devices ($f=10\times8=80$ bundles) to simulate 100 devices ($N=100$) for calculation. Furthermore, we need to acquire the parameters related to runtime through empirical values or pre-experimental measurements before dynamic allocation. These parameters include the average duration for the completion of the scheduled task in \emph{Logical Simulation} with $c$ grades of devices, denoted as $\{\alpha_1, \alpha_2, \ldots, \alpha_c\}$, the corresponding average duration $\{\beta_1, \beta_2, \ldots, \beta_c\}$ and the startup time of compute framework $\{\lambda_1, \lambda_2, \ldots, \lambda_c\}$ on physical phone devices.

To optimize the allocation strategy, we suppose $\{x_1, x_2, \ldots, x_c\}$ simulated devices are allocated to the \emph{Logical Simulation} and $\{N_1-q_1-x_1, N_2-q_2-x_2, \ldots, N_c-q_c-x_c\}$ to \emph{Device Simulation}. Note that the single unit resource bundle in \emph{Logical Simulation} and the single physical device in \emph{Device Simulation} are both capable of repetitive emulation of multiple devices computation, and all resources support parallel processing. Considering the above, the following can be deduced: the task duration of \emph{Logical Simulation} is given by $T_l=max\{\lceil\frac{k_ix_i}{f_i}\rceil\alpha_i\}, i=1,\ldots,c$, the task duration of \emph{Device Simulation} is expressed as $T_p=max\{ \lceil\frac{N_i-q_i-x_i}{m_i}\rceil\beta_i+\lambda_i\}, i=1,\ldots,c$. Consequently, the total duration to execute the task is $T=\max\{T_l, T_p\}$.

The problem of minimizing $T$ can be formulated as a integer linear programming problem, subject to the conditions:
\begin{equation}
\left\{  
\begin{array}{ll}  
    T-\lceil\frac{k_ix_i}{f_i}\rceil\alpha_i\geq 0 \\[5pt] 
    T-\lceil\frac{N_i-q_i-x_i}{m_i}\rceil\beta_i-\lambda_i\geq 0 \\  
\end{array}  
\right.
, 0\leq x_i \leq N_i-q_i.
\label{eq}\end{equation}
Furthermore, if it is necessary to impose constraints on resource usage during allocation, such as prioritizing the use of \emph{Logical Simulation} resources, after obtaining the minimum value of $T$, then we can find the combination from all feasible solutions that maximizes $\sum_{i=1}^{c}{x_i}$. 

By solving this linear programming problem and allocating the respective quantities of simulation devices across resources, the system facilitates the execution of tasks using hybrid heterogeneous resources. This approach ensures the acquisition of precise metrics from physical devices while maintaining low temporal costs.
\begin{figure}[!b]
  \centering
  \includegraphics[width=\linewidth]{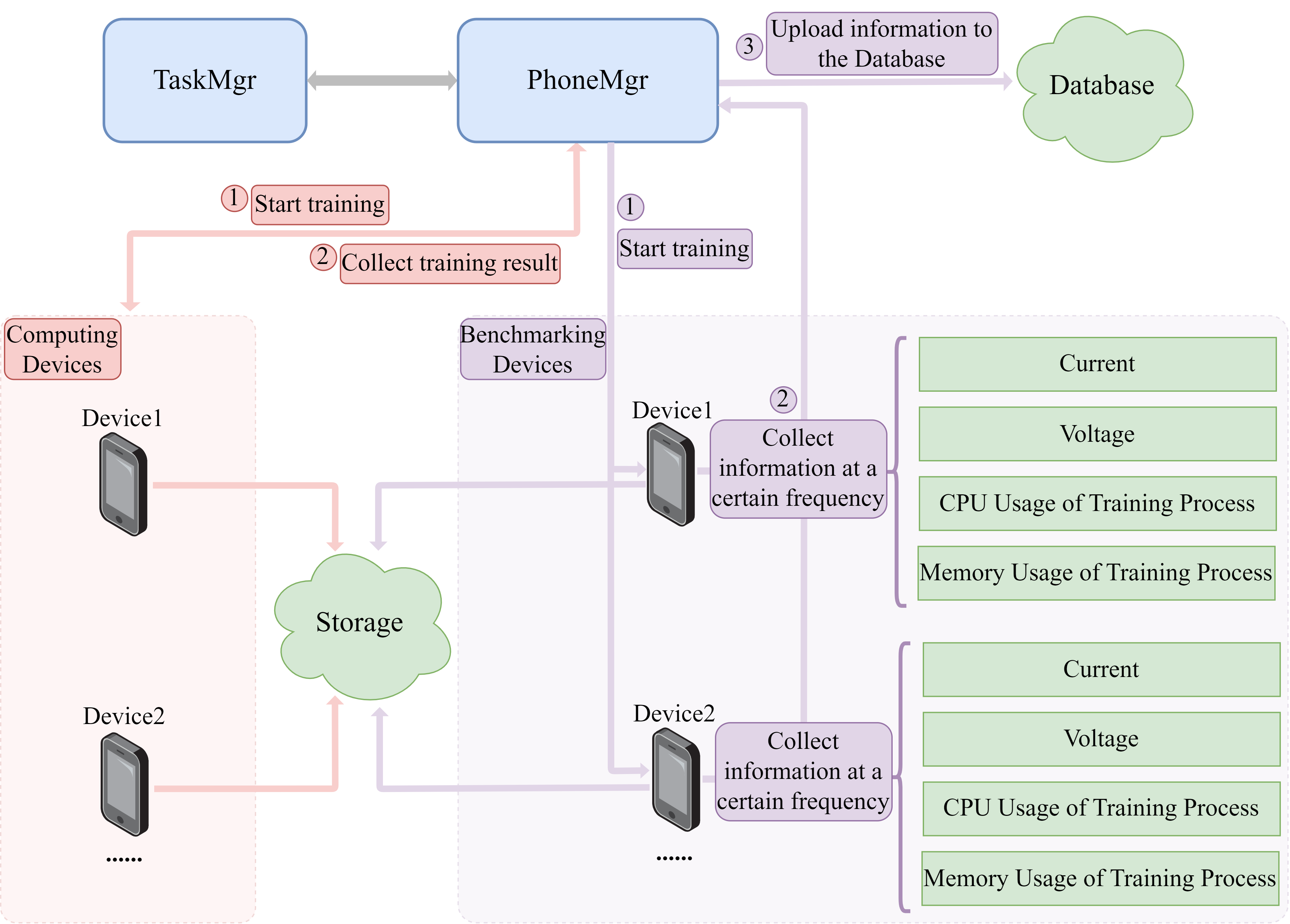}
  \caption{\emph{PhoneMgr} manages the physical devices cluster and performance measurement.}
  \label{fig:measure_performance}
\end{figure}

\subsection{Physical Devices Performance Measurement}

SimDC implements a device management module called \emph{PhoneMgr}, designed for efficient management of a physical devices cluster. \emph{PhoneMgr} performs various operations and interface management for physical devices, primarily relying on ADB commands, and supports user-defined ADB scripts for phones running third-party applications (APKs). ADB is a versatile command-line tool capable of communicating with Android devices, executing numerous device actions such as installing and debugging applications, and providing access to the Unix shell. This client-server architecture tool effectively enhances the flexibility and control of device management.

To evaluate power consumption and performance during the executing process, the physical device cluster is configured with devices serving as \emph{Computing Devices}, and specialized devices integrated to measure performance known as \emph{Benchmarking Devices}. When tasks include settings for physical devices' performance measurement, \emph{PhoneMgr} distinguishes between these two types of devices. Once the Benchmarking devices start training, \emph{PhoneMgr} retrieves information from these devices at a certain frequency, organizes it in real-time, and uploads it to the cloud database for storage, as shown in Fig. \ref{fig:measure_performance}. The collected basic device information typically includes current (in $\mu$A), device voltage (in mV), CPU usage ($\%$), memory usage (in KB), and bandwidth usage (in B). Specific retrieval methods are as follows:
\begin{itemize}
\item Current ($\mu$A):

\emph{adb shell cat /sys/class/power\_supply/battery/current\_now}

\item Voltage (mV):

\emph{adb shell cat /sys/class/power\_supply/battery/voltage\_now}

\item CPU usage ($\%$):

\emph{adb shell top -b -n 1 -p $<$process\_pid$>$}, where process\_pid is the process pid of the executing process, obtainable via \emph{adb shell pgrep -f $<$process\_name$>$}, with $<$process\_name$>$ being the name of the process (i.e., the name of the training Android package).

\item Memory usage (K):

\emph{adb shell dumpsys $<$process\_name$>$$|$grep PSS}

\item Bandwidth usage (B):

\emph{adb shell cat /proc/$<$process\_pid$>$/net/dev $|$grep wlan}, which encompasses both received and transmitted data that need to be extracted and summed.
\end{itemize}
The information collected typically contains other non-essential data, requiring post-processing to extract valid data. Once the Benchmarking Devices complete executing, \emph{PhoneMgr} will stop information collection, thereby concluding the measurement of the physical devices' performance.

\section{Device Behavior Traffic Controller}

\begin{figure}[!b]
  \centering
  \includegraphics[width=\linewidth]{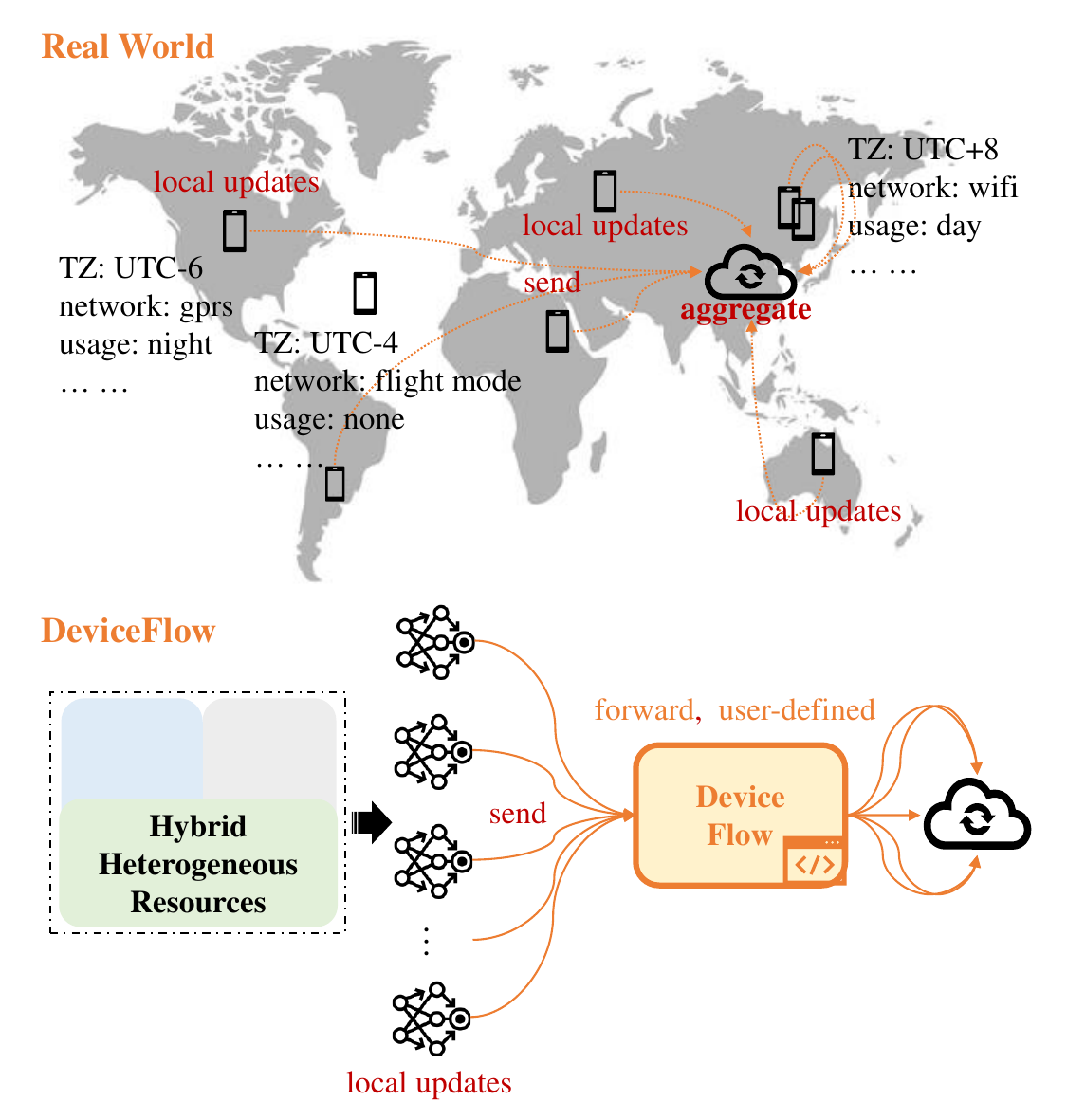}
  \caption{Real-world applications and \emph{DeviceFlow} in device-cloud collaboration.}
  \label{fig:deviceflow_background}
\end{figure}

\subsection{DeviceFlow Architecture}

In real-world practical applications, as illustrated in Fig. \ref{fig:deviceflow_background}, physical phones often exhibit diverse behaviors due to factors such as timezones, environmental networks, user actions, and inherent variability. The behavior of these large-scale devices can significantly influence the computational performance of cloud services in device-cloud collaboration scenarios, potentially leading to issues such as cloud computing (e.g., model agg)  bias\cite{zhou2022role, rodio2023federated}. To accurately simulate the operation of large-scale devices in such scenarios, we propose \emph{DeviceFlow}, a device behavior traffic controller designed to emulate the interaction traffic between devices and cloud services.

When edge devices collaborate with cloud services, they typically upload computation results to storage upon task completion and transmit messages to cloud services. Cloud services then retrieve the corresponding data from storage based on the received messages for further processing. \emph{DeviceFlow} operates as an intermediary component, bridging edge devices and cloud services by managing message transmission. From the perspective of edge devices, \emph{DeviceFlow} functions as a proxy for the cloud, while from the viewpoint of cloud services, it serves as a representation of the edge devices.

\begin{figure}[t]
  \centering
  \includegraphics[width=\linewidth]{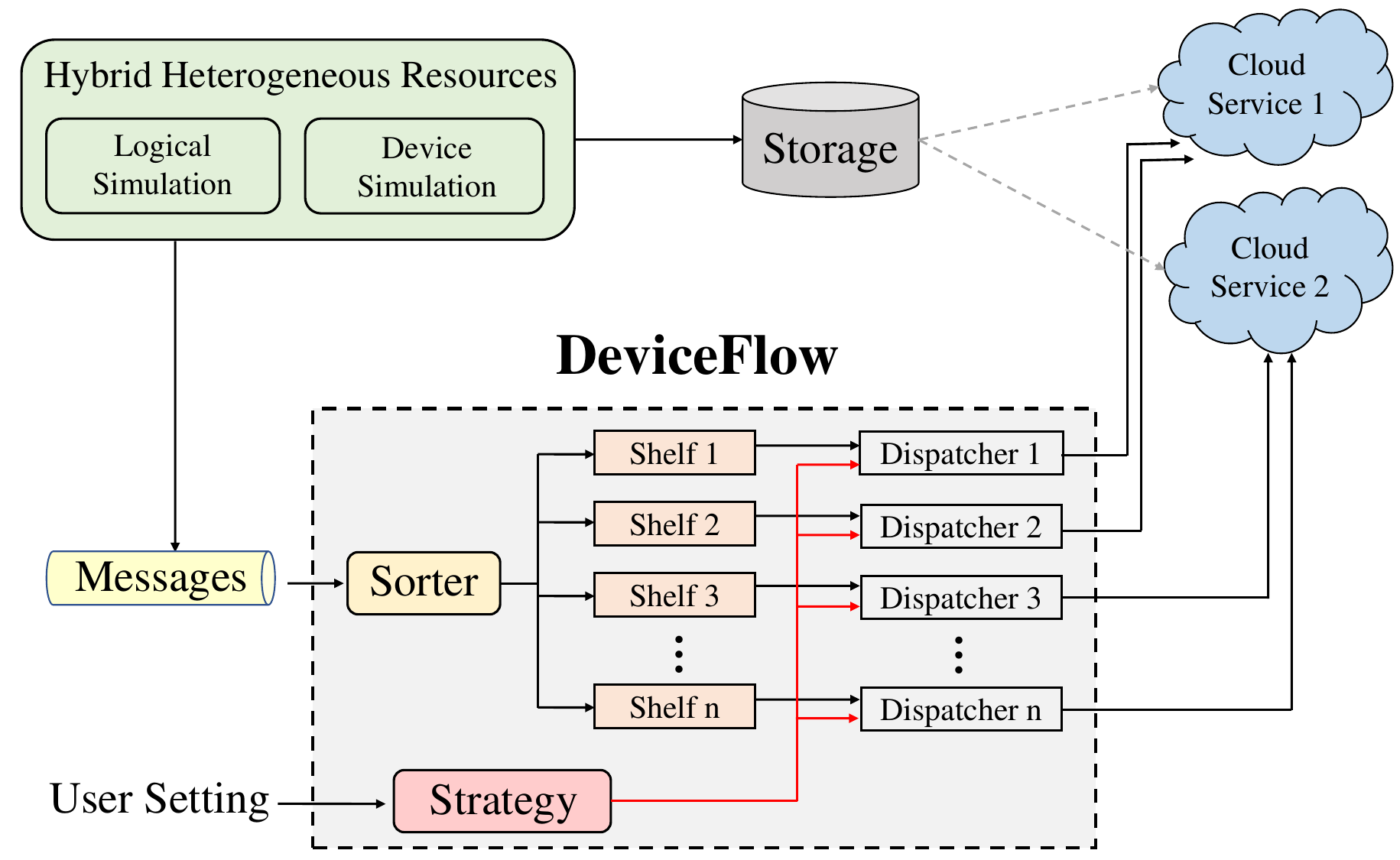}
  \caption{The framework of \emph{DeviceFlow}.}
  \label{fig:deviceflow_framework}
\end{figure}

The architecture of \emph{DeviceFlow}, as depicted in Fig. \ref{fig:deviceflow_framework}, comprises four core modules: Sorter, Shelf, Dispatcher, and Strategy. When task submission involves the use of \emph{DeviceFlow}, the Strategy module stores user-defined strategies for message dispatching. The Sorter module is responsible for receiving messages from computational clusters and determining the appropriate Shelf for storage based on the task\_id within the messages. During the initiation and completion of each round of operator flow within a task, computational clusters send messages from logical or device simulation to \emph{DeviceFlow}. Upon receiving these messages, \emph{DeviceFlow} activates the Dispatcher module which handles the message dispatching. The Dispatcher module first retrieves and parses the corresponding strategy from the Strategy module, then extracts the pending messages from the Shelf module and dispatches them to the cloud services according to the predefined strategy. The Dispatcher modules associated with different Shelf modules operate independently, ensuring that the dispatch processes of different tasks remain isolated and do not interfere.

\subsection{Message Dispatching Strategy}

In \emph{DeviceFlow}, users can define various strategies to control message dispatching and simulate device dropouts, thereby enabling high-fidelity and flexible simulation of device behaviors. The dispatching strategies are categorized into real-time accumulated dispatching and rule-based dispatching.

The real-time accumulated dispatching strategy is activated at the beginning of each round within the task. Once the accumulated number of messages received by \emph{DeviceFlow} reaches a predefined threshold (n messages), these messages are immediately dispatched to downstream cloud services. The value of n can be configured within the range of 1 to the system's maximum capacity. When n is set to 1, \emph{DeviceFlow} operates as a real-time transmission mechanism. To more accurately emulate the occurrence of device disconnections in real-world scenarios, the strategy incorporates a probability of single message transmission failure, denoted as 
$p\in[0,1]$. By simulating device dropouts, this strategy replicates the inherent unpredictability of network conditions during device-cloud collaboration, thereby enhancing the fidelity of the simulation.

The rule-based dispatching strategy is initiated upon the completion of each round within the task. Guided by user-defined dispatching rules, this strategy transmits accumulated messages to cloud-side services at specific time points or during designated time intervals. The specific time-point dispatching mechanism operates based on user-defined time points for sending messages, which may include relative times (measured from the end of the round) or absolute times, and dispatches messages according to the user-specified quantity. In this mechanism, \emph{DeviceFlow} offers two methods for simulating dropouts: the probability of transmission failure can be set for each time point, and a random selection of a certain number of messages can be discarded at each time point.

The specific time-interval dispatching mechanism enables users to define a custom transmission rate function within a designated time interval, thereby conveniently simulating real-world devices behavior traffic. In this mechanism, the independent variable is time $t$, and the dependent variable is the transmission rate $y=f(t)$. First, the total amount of pending messages is equated to the total area under the curve (AUC) $y=f(t)$ over its entire domain. Second, based on the single-threaded transmission capacity limit of \emph{DeviceFlow} (e.g., 700 messages per second), a reasonable discrete transmission time interval is calculated and determined to ensure that the number of messages sent at any single point does not exceed the transmission capacity limit and that the interval is sufficiently small. Finally, the corresponding dispatching quantity is calculated for each discrete interval based on the AUC ratios with total AUC, and the starting point of the interval is taken as the transmission time point. These above operations transform the specific time-interval dispatching mechanism into the aforementioned specific time-point dispatching mechanism for execution. It is important to note that the domain of $t$ is a closed interval, which can be scaled to align with the user-defined specific time interval. The transmission rate function $y$ must be a single-valued, bounded, non-negative continuous function, supporting piecewise continuity. The user-defined specific time interval also supports both relative and absolute time settings. 
This mechanism also supports two dropout methods: probability and random discard in each time interval.
\section{Experiments}\label{sec:experiment}
We conduct specific experiments to assess the efficacy of our proposed SimDC platform. 

\subsection{Experimental Settings}\label{exp:settings}
\subsubsection{Dataset and Model}
Device-cloud collaboration deployment is one of the most common scenarios in recommendation systems. Within this study, we focus on one of the crucial problems in this area: click-through rate (CTR) prediction. We use the public Avazu dataset, which contains a vast amount of advertising click data, including features such as device information, application and website classification, and anonymized classification variables. For our experiments, we select a subset comprising approximately 2 million records, covering 100,000 unique devices identified by device\_id as the training set, and we additionally select records from 1,000 unique devices as the test set.

Considering factors such as training duration, computational power, and power consumption on physical devices, the industry currently favors simpler and more efficient models for CTR prediction in edge-cloud scenarios. Model like logistic regression (LR) is particularly suitable for large-scale data and real-time prediction tasks due to their simplicity and efficiency. Therefore, we employ the LR model as a benchmark to verify the practicality of the proposed platform in real-world application scenarios. For model aggregation, we adopt FedAvg, with learning rate set to 1e-3 and the number of local training epochs on the client side set to 10. We evaluate the experimental results using the model from each aggregation.

\subsubsection{Experimental Environment}
We construct two heterogeneous resource clusters for simulation experiments: a Ray cluster and a physical devices cluster. The Ray cluster is used for large-scale logical simulations, with a default configuration of 200 CPU cores and 300 GB memory, supporting elastic scaling. When utilizing the Ray cluster, each task creates a placement group, where each actor corresponds to a resource bundle, simulating different grades of devices. Further, we categorize the simulated devices into two grades: High devices with a default configuration of 4 CPU cores and 12 GB memory, and Low devices with 1 CPU core and 6 GB memory. The resources configured to each actor can be adjusted based on actual needs to meet different experimental requirements.

In the physical devices cluster, we have a default configuration of 10 local physical devices and 20 remote MSP devices. Similar to the logical simulation, the physical devices are divided into High (4 devices, with more than 8 GB memory) and Low (6 devices, with less than 8 GB memory) grades. MSP devices are also categorized into High (13 devices) and Low (7 devices) grades. Physical devices can be further classified based on device models, CPU frequency, NPU support, etc., while the logical simulation can mirror such classifications by setting up resource bundles for the simulated devices.

Since current efforts to simulate large-scale device computations are mainly focused on federated learning frameworks, we limit to comparing the effectiveness of SimDC in large-scale device simulations with several mainstream federated learning frameworks, including FedScale\cite{lai2022fedscale} and FederatedScope\cite{federatedscope} (FS-Real\cite{fsrealvldb,fsrealkdd} is one version of FederatedScope). It is important to note that SimDC fundamentally differs from these federated learning frameworks. SimDC is a device simulation platform designed for device-cloud collaborative scenarios, capable of simulating edge-side training, inference, specific calculations, and collective behavior traffic, while federated learning is just one of its applications. In related experiments, we focus solely on comparing the performance of these methods in large-scale device simulation calculations.    
\subsection{Hybrid Heterogeneous Resources}\label{expsec:hybrid}

\subsubsection{Simultaneous Large-Scale Simulations and Physical Device Performance Measurement} 

Tasks can request and utilize hybrid heterogeneous resources for computation, enabling the collection of performance metrics from physical devices within the clusters while conducting large-scale logical simulations. In our experiment, we simulate 500 High and 500 Low grade devices. Additionally, we configure 5 physical devices per grade as benchmarking devices exclusively for training and performance measurement. These devices are not reused as computation units in a single round for simulating device computations. 

\begin{table}[htbp]
  \centering
  \caption{Measurement of physical performance metrics during simulation.}  
  \label{tab:performance}
  \resizebox{\linewidth}{!}{  
    \begin{tabular}{ccccc}
      \toprule  
      Grade & Stage & Power (mAh) & Duration (min) & Commu(KB)\\  
      \midrule  
              & 1 no APK initiated & 0.24 & 0.25 &     \\  
              & 2 APK launch& 0.51 & 0.25 &     \\  
        High  & 3 Training  & 0.18 & 0.27 & 33.10 \\  
              & 4 Post-training & 0.37 & 0.25 &     \\  
              & 5 Closure of APK& 0.44 & 0.25 &     \\
      \midrule 
              & 1 no APK initiated & 1.71 & 0.25 &     \\  
              & 2 APK launch & 1.80 & 0.25 &     \\  
        Low   & 3 Training & 0.66 & 0.36 & 33.10 \\  
              & 4 Post-training & 1.65 & 0.25 &     \\  
              & 5 Closure of APK & 1.82 & 0.25 &     \\      
      \bottomrule  
    \end{tabular}}   
\end{table}

\begin{figure}[t]
  \centering
  \includegraphics[width=\linewidth]{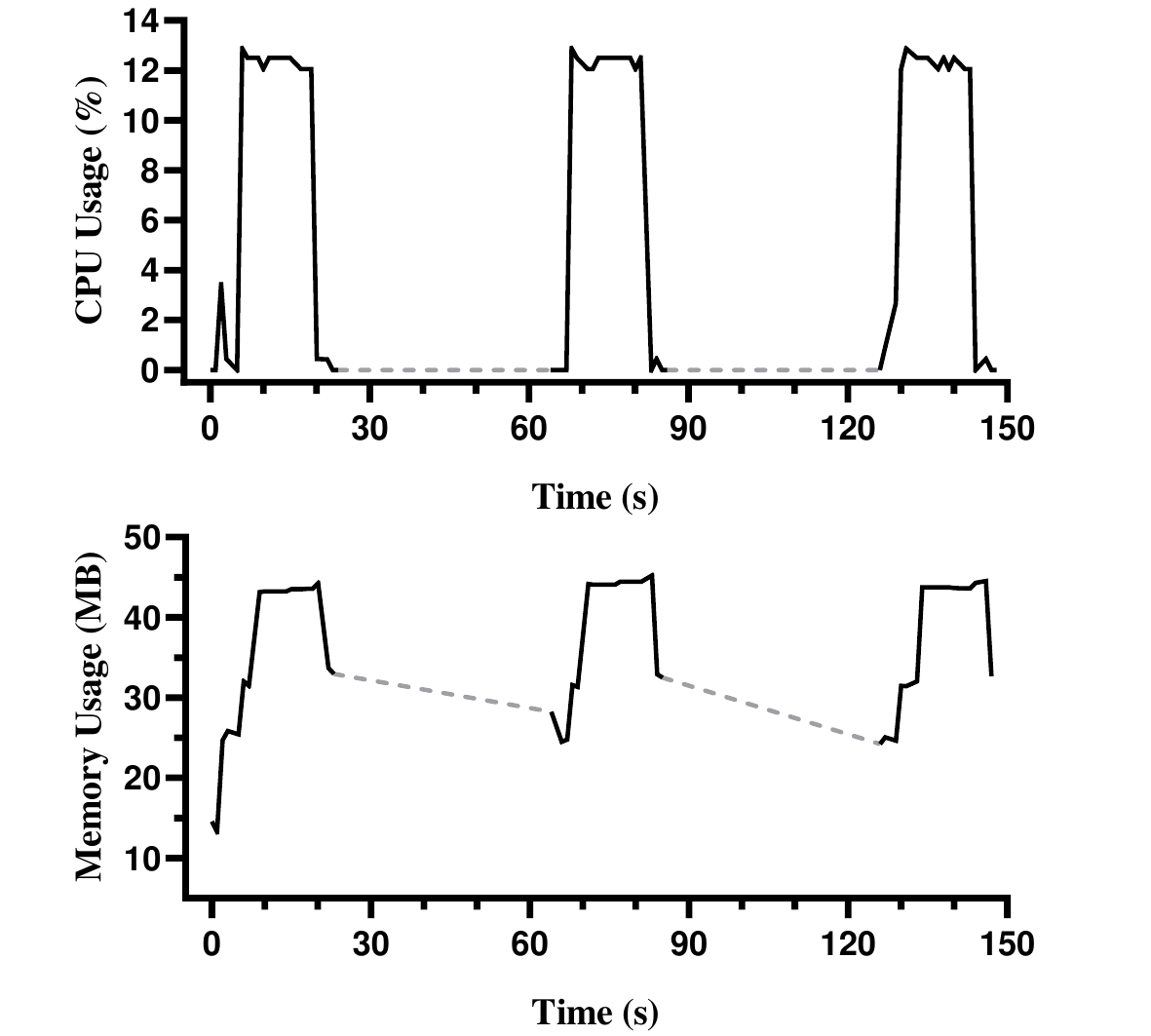}
  \caption{Measurement of CPU and memory usage during the first three rounds.}
  \label{fig:devices_cpu_and_mem}
\end{figure}

Table \ref{tab:performance} presents several physical performance metrics recorded during the initial training round for benchmarking devices, including average power consumption, runtime, and communication volume across five different stages: Stage 1 - clearing background tasks without running the APK; Stage 2 - launching the APK without starting training; Stage 3 - training using the APK; Stage 4 - post-training with the APK still active; and Stage 5 - exiting the APK and clearing background tasks. Notably, under identical tasks, High-grade devices exhibit shorter runtime and lower power consumption compared to Low-grade devices. Fig. \ref{fig:devices_cpu_and_mem} illustrates the CPU and memory usage of a specific benchmarking device during the first three training rounds, as denoted by the solid black line. Performance measurement starts with the APK launch, and no data is recorded during the device's wait for global aggregation to complete (the dashed gray lines). These comprehensive metrics offer valuable insights for optimizing edge-side algorithms and developing cloud-side configuration strategies.

\subsubsection{Task Accuracy Unaffected by Hybrid Heterogeneous Computing}
Client-side federated learning algorithms are typically integrated into specific business APKs that must adhere to requirements such as power consumption and size upon deployment. Given the cost constraints and implementation complexities, it is nearly impossible to embed a uniform federated learning framework across different APKs, due to inevitable variations in device-cloud communication components and shared storage requirements. Logical and device simulations often utilize operators with identical functionalities but differing underlying implementations. Additionally, disparities in hardware architecture and compilation optimizations between servers and mobile devices can lead to variations when executing the same operator across platforms. Therefore, it is crucial to evaluate the potential impact of hybrid heterogeneous resource computing on algorithm accuracy.

\begin{figure*}[tbp]
\begin{minipage}[t]{0.33\textwidth}
\centering
  \includegraphics[height=3.4cm]{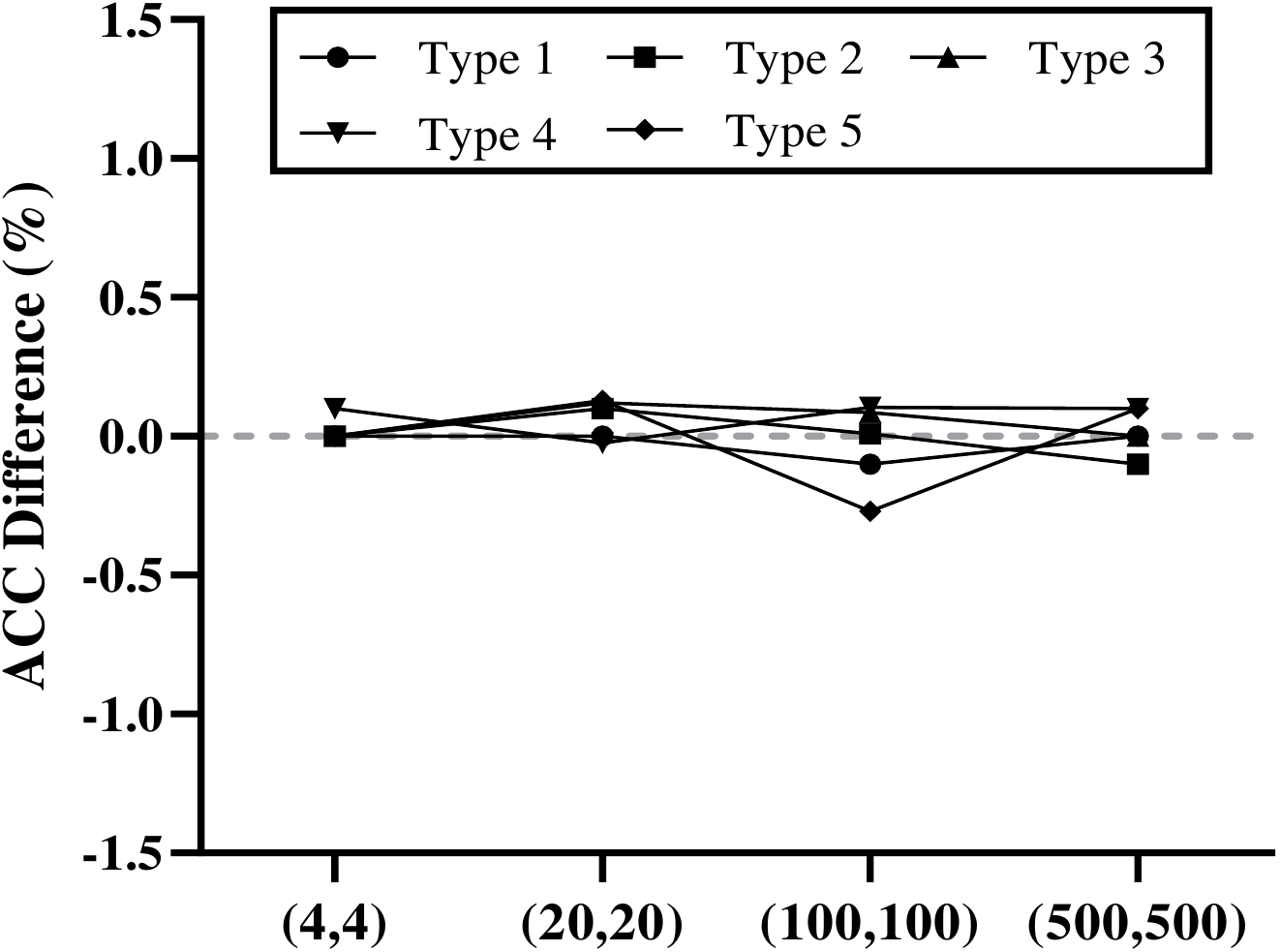}
  \caption{Accuracy difference relative to scale in two grades of devices.}
  \label{fig:acc_diff}
\end{minipage}
\begin{minipage}[t]{0.33\textwidth}
\centering
  \includegraphics[height=3.4cm]{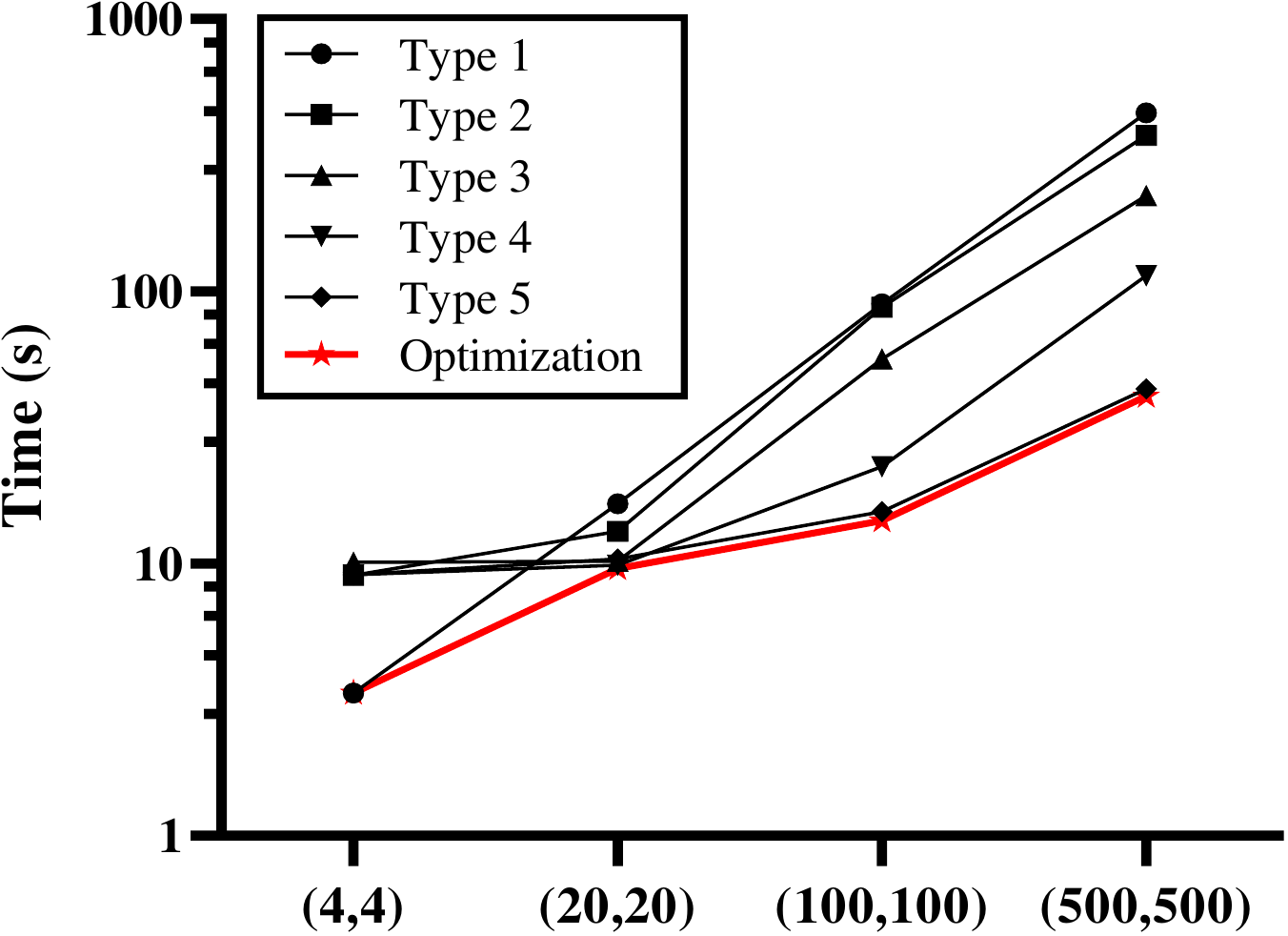}
  \caption{Execution times vs. scale.}
  \label{fig:time_scale}
\end{minipage}
\begin{minipage}[t]{0.33\textwidth}
\centering
  \includegraphics[height=3.4cm]{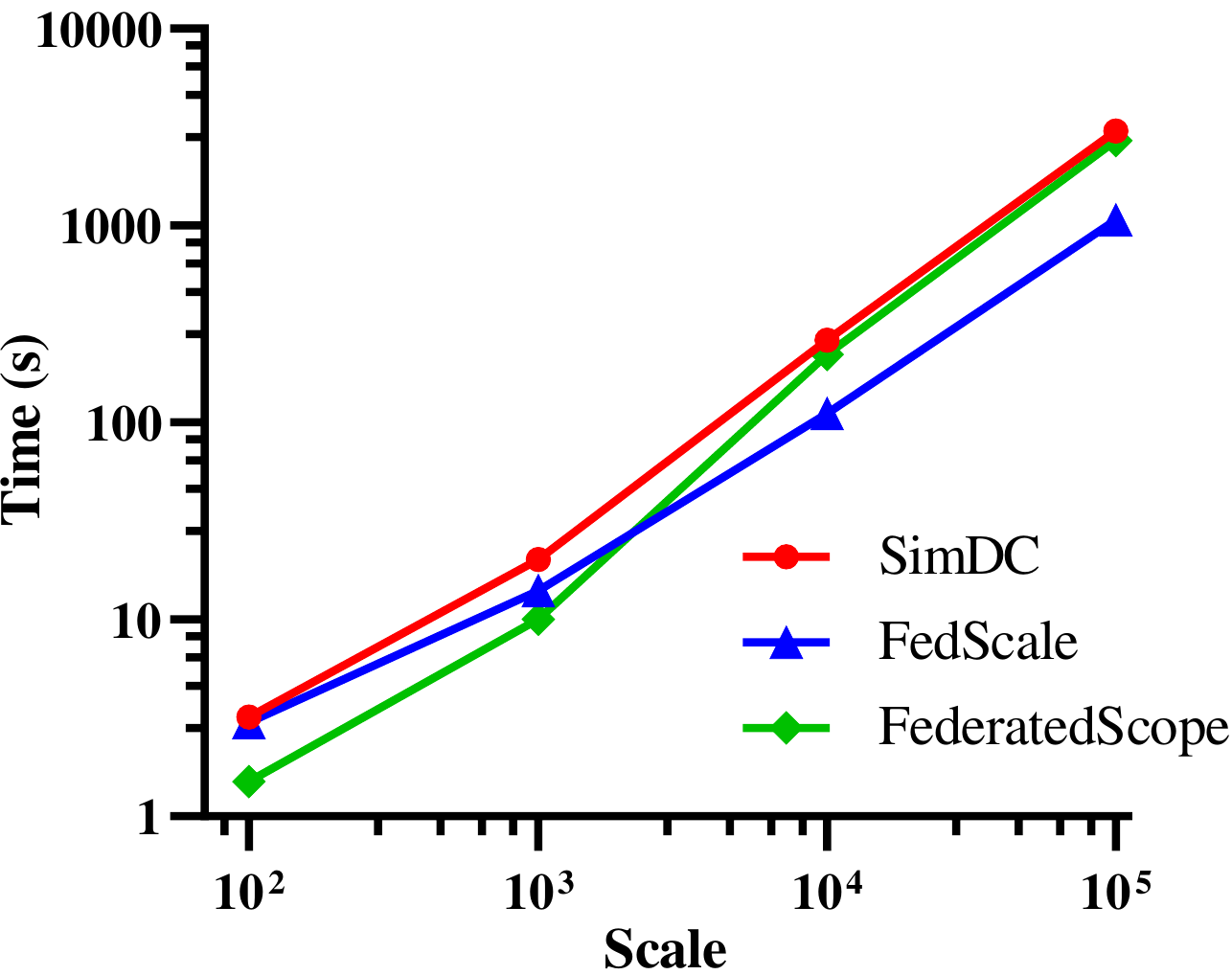}
  \caption{Scalability of popular simulators.}
  \label{fig:scalability_fig}
\end{minipage}
\end{figure*}

In our implementation, the training operators used in logical simulation are based on the PyMNN architecture, while device simulation employs operators from the C++ MNN architecture used in actual business SDKs. The experiment simulates two grades of devices and performs $10$ rounds of device-cloud co-computing. To consider the potential effects of hybrid heterogeneous configurations on training outcomes, we establish five representative simulation allocation ratios, expressed as (\emph{Logical Simulation}, \emph{Device Simulation}): \emph{Type $1$}: ($100\%$, $0\%$), \emph{Type $2$}: ($75\%$, $25\%$), \emph{Type $3$}: ($50\%$, $50\%$), \emph{Type $4$}: ($25\%$, $75\%$), \emph{Type $5$}: ($0\%$, $100\%$). For each grade of device (High, Low), we simulate scalable scenarios with $4$, $20$, $100$, and $500$ devices, and compare the accuracy (ACC) difference between the hybrid heterogeneous settings and the benchmark local distributed computing environments. As shown in Fig. \ref{fig:acc_diff}, across different device scales and allocation ratios, the ACC differences were consistently below 0.5$\%$, within the acceptable margin of error. The result indicates that hybrid heterogeneous resource computing does not significantly impact algorithm accuracy for device-cloud simulations.

\subsubsection{Optimized Device Allocation Minimizes Task Execution Time}
In the aforementioned scenarios of varying scales, we also compared the single task execution time between our proposed hybrid allocation optimization and the five typical allocation ratios. As shown in Fig. \ref{fig:time_scale}, for smaller scales, execution time on physical devices is primarily influenced by the APK startup time, making logical simulation relatively faster. In contrast, at larger scales, the execution time per round is mainly determined by the training time itself, where the underlying implementation of device simulation operators executes faster. The red line indicates the task execution time using our optimization algorithm at different scales. It consistently demonstrates shorter execution time compared to other allocation ratios, validating the effectiveness of the proposed optimization approach.



\subsubsection{Scalability}
Lastly, we validated the scalability of the SimDC platform. Using a server cluster with 200 CPU cores, we simulated edge devices of a single performance level, scaling from 100 to 100,000 devices. Fig. \ref{fig:scalability_fig} depicts the average single-round training times of SimDC, FedScale, and FederatedScope across different scales. The results show that for fewer than 1,000 devices, the single-round training time of SimDC is larger than that of the other two frameworks. Notably, FedScale does not use device-cloud communication during simulations. Its data and models are stored directly in memory, and data is transferred only between memories when simulating different clients. FederatedScope employs a similar strategy for data and models and can only use a single resource instance to simulate clients. In contrast, SimDC's logical simulation based on the Ray framework supports the distribution of resources across different physical servers. In this setup, each actor in the logical simulation must download the corresponding data and model for its simulated devices, and device simulation require each physical device to download the relevant data and model during computation for corresponding simulated devices. These results then be sent to shared storage and communicated with cloud services. As a result, although SimDC takes longer for fewer devices, its architecture more closely mirrors real-world business applications. 

Additionally, both SimDC and FederatedScope independently simulate clients and use device-cloud communication for aggregation. Furthermore, as the number of simulated devices increases, especially beyond 10,000, the single-round execution time is primarily dominated by the device scale. The single-round training times of SimDC and FederatedScope are comparable at large scales, indicating SimDC’s capability for large-scale simulations. While FedScale appears faster, its simulation deviate significantly from real-world scenarios.

\subsection{Device Behavior Traffic Simulation}\label{expsec:deviceflow}

\subsubsection{Impact of Device Behavior Traffic}
Verifying the direct impact of different device behavior traffic is important. Taking federated learning as an example, large-scale business scenarios typically involve edge devices from various time zones and regions. These devices exhibit response delays during training and have varied data distributions due to differences in user behavior. The traffic received by the cloud service from these edge devices during execution is a typical representation of this phenomenon. In non-independent and identically distributed (non-IID) scenarios, different group traffic curves can significantly impact cloud computation results, such as aggregation performance. Therefore, the cloud service may need to fine-tune hyperparameters based on the anticipated traffic curves to achieve optimal outcomes under realistic conditions.

In real federated learning scenarios, the cloud usually does not know the exact number of participating devices or samples per training round in advance. Therefore, conditions must be set to trigger aggregation. Common triggers include reaching a threshold of total edge training samples or reaching scheduled times. We constructed a non-IID scenario where clients with higher CTR transmit data faster to the cloud, while those with lower CTR experience longer delays. In this scenario, we designed three typical response curves using right-tailed normal distributions ($\mathcal N(0, \sigma)$) with $\sigma$ values of 1, 2, and 3.

\begin{figure}[b]
  \centering
  \includegraphics[width=\linewidth]{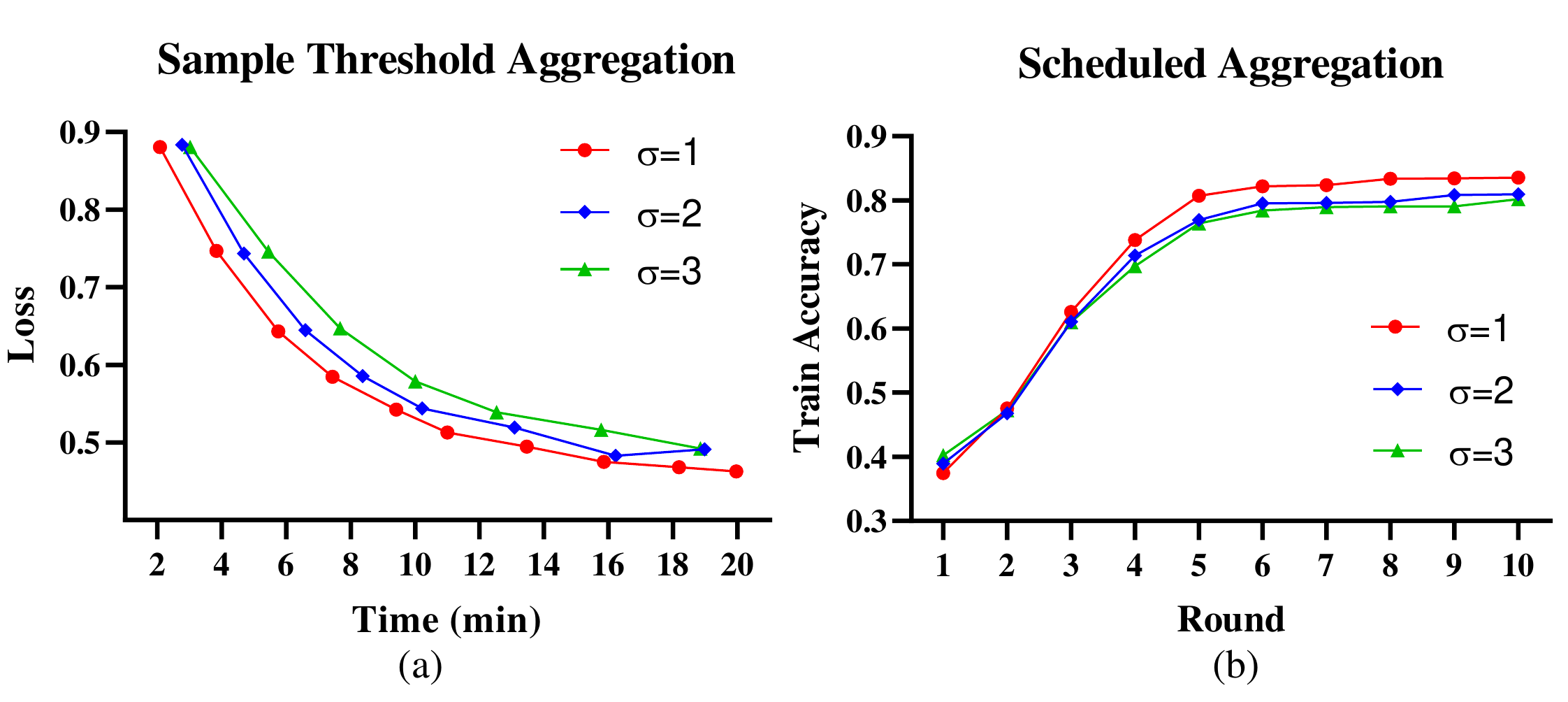}
  \caption{Impact of device behavior traffic curves on aggregations.}
  \label{fig:deviceflow_diffagg}
\end{figure}

\begin{figure*}[t]
  \centering
  \includegraphics[height=3.23cm]{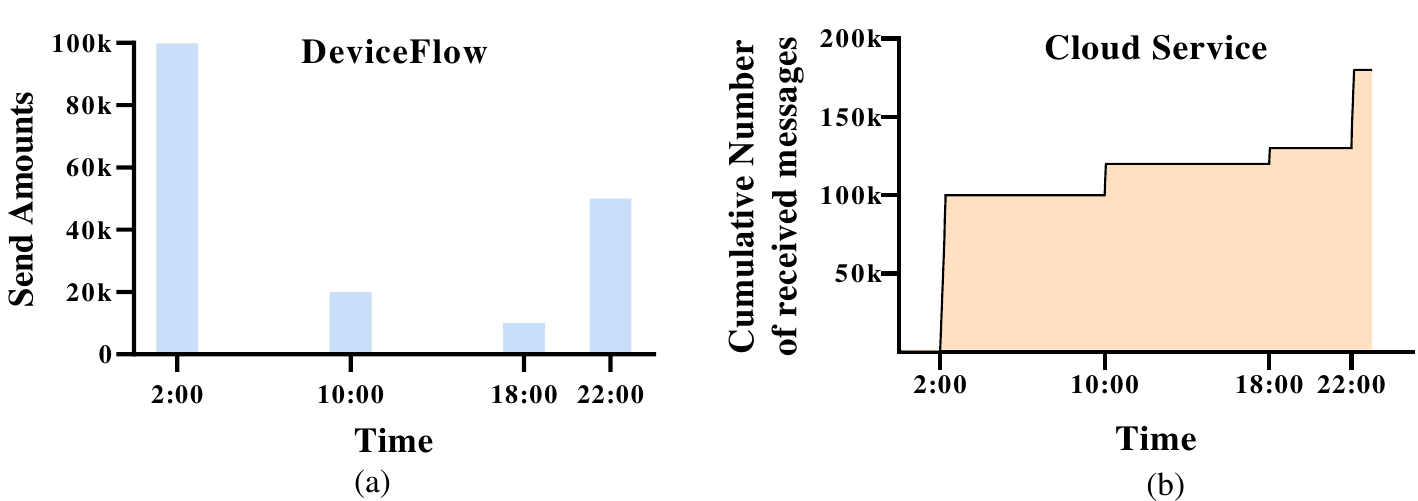}
  \includegraphics[height=3.23cm]{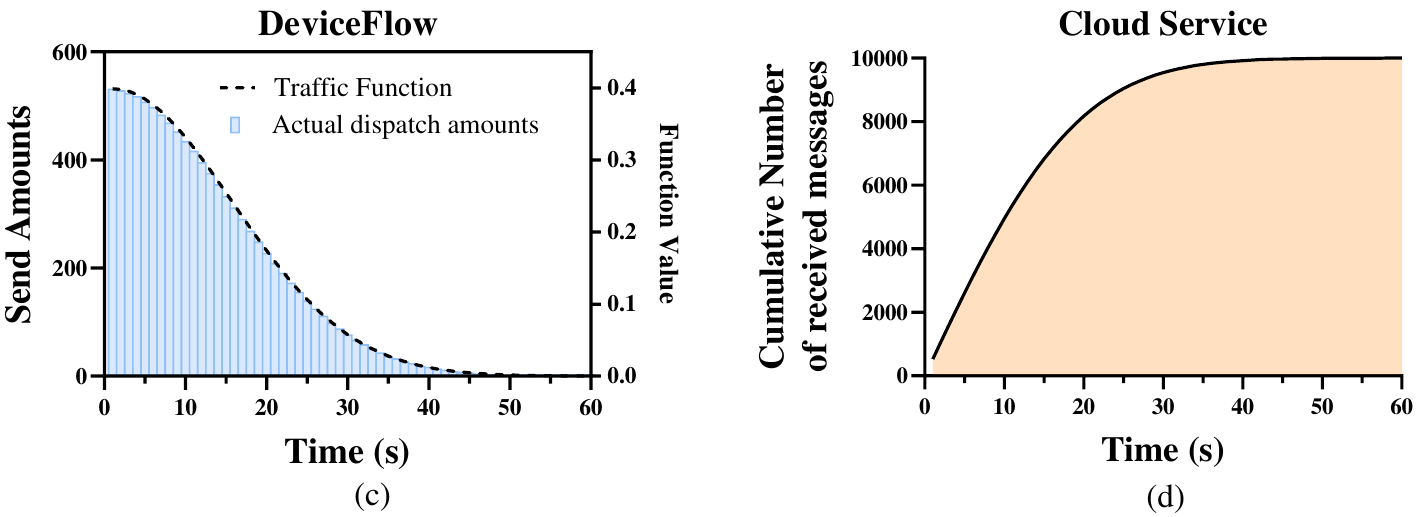}
  \caption{Rule-base dispatch strategies.}
  \label{fig:deviceflow_rule_based}
\end{figure*}

Under the "sample threshold" condition, a smaller $\sigma$ in the traffic curve indicates that a large number of samples reach the aggregation service in a shorter time, thereby shortening the aggregation time for each round. As shown in Fig. \ref{fig:deviceflow_diffagg}(a), within a fixed 20-minute window, the red curve ($\sigma=1$) completes more aggregation rounds, resulting in lower loss. Under the "scheduled aggregation" condition, a smaller $\sigma$ in the traffic curve allows more samples to be aggregated within the fixed time, making the aggregated dataset more representative of the true distribution. Fig. \ref{fig:deviceflow_diffagg}(b) shows that the red curve ($\sigma=1$) achieves higher train accuracy per round compared to the other curves. These results demonstrate that different device behavior traffic curves can significantly impact the computation results of cloud services. Consequently, cloud services can optimize strategies or fine-tune hyperparameters based on the observed device behavior traffic curves.

\subsubsection{Deviceflow Dispatch Strategies}
We categorize device behavior patterns into two types: real-time accumulated dispatching and rule-based dispatching. The real-time accumulated dispatching strategy sends messages based on a user-defined sequence. For instance, if the user sets this strategy to [20, 100, 50], \emph{DeviceFlow} will cycle through sending a certain amount of messages to the downstream cloud service according to this sequence until the simulation task is completed. When the real-time accumulated dispatching is set to [1], it behaves like other simulators, immediately sending messages to the cloud service after computation.

Rule-based dispatching includes two mechanisms: specific time-point dispatching and specific time-interval dispatching. These mechanisms enable a more precise simulation of user behavior. As depicted in Fig. \ref{fig:deviceflow_rule_based}(a), specific time-point dispatching dictates certain transmission amounts at distinct time points. The cloud service then receives the corresponding volume of messages, with the cumulative number of messages illustrated in Fig. \ref{fig:deviceflow_rule_based}(b). Note that the rate of message dispatching is subject to an upper limit, the cloud service actually receives the full messages over a period spanning the designated time point and subsequent certain intervals.

\begin{table}[b]
  \centering
  \caption{Similarity between user-defined traffic curves with \emph{Deviceflow} actual dispatch strategies.}  
  \label{tab:correlation}
    \begin{tabular}{ccccc}
      \toprule  
      User-defined traffic curves & Domain & Correlation coefficient\\  
      \midrule  
      $\mathcal N$(0, 1) & [-4, 4]      & 0.999 \\  
      $\mathcal N$(0, 2) & [-4, 4]      & 0.999 \\  
      sin(t)+1           & [0, 6$\pi$]  & 0.995 \\  
      cos(t)+1           & [0, 6$\pi$]  & 0.996 \\  
      $2^t$              & [0, 3]       & 0.999 \\
      $10^t$             & [0, 3]       & 0.999 \\
      \bottomrule  
    \end{tabular}
\end{table}

A specific time-interval dispatching strategy is represented by the dashed line in Fig. \ref{fig:deviceflow_rule_based}(c), where the user defines a right-tailed normal distribution traffic curve with $\sigma$=1. The domain and range of the function are scaled proportionally based on the actual dispatch interval (1 min) and volume (10000). Fig. \ref{fig:deviceflow_rule_based}(c) shows the discretized per-second sending volumes, while Fig. \ref{fig:deviceflow_rule_based})(d) displays the cumulative messages received by the cloud service. It is evident that \emph{DeviceFlow}'s actual dispatch strategy closely matches the user-defined traffic curve. We further compared the similarity between \emph{DeviceFlow}'s actual dispatch strategy and the user-defined traffic curves for various single-value bounded non-negative continuous functions. As shown in Table \ref{tab:correlation}, the Pearson correlation coefficients exceed 0.99 in all cases, confirming that \emph{DeviceFlow} can accurately simulate device behavior traffic curves, whether they are continuous, periodic, or exhibit significant variability in range.

\begin{figure}[t]
  \centering
  \includegraphics[width=\linewidth]{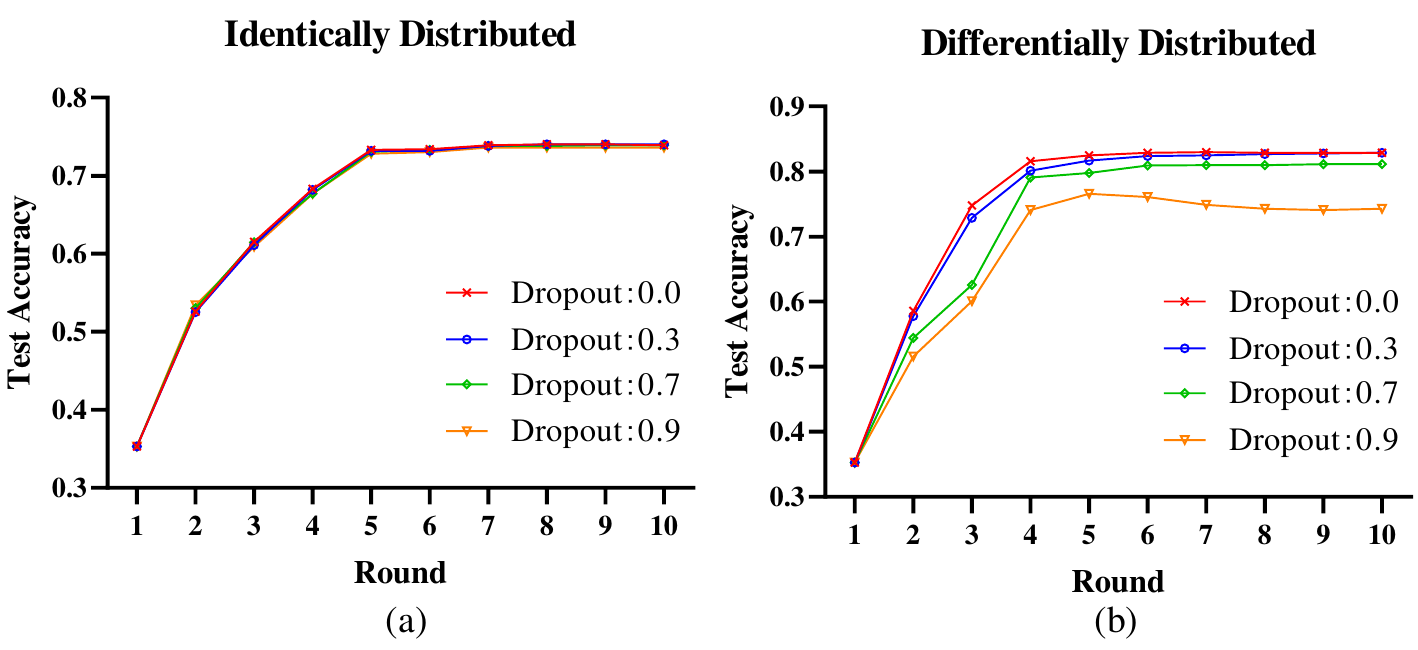}
  \caption{The impact of device dropout on different data distribution.}
  \label{fig:deviceflow_dropout}
\end{figure}

Finally, we examined the impact of device dropout on the computation results of the cloud service. In the real-time dispatching scenario, we simulated 1,000 devices with varying dropout probabilities (0.3, 0.7, 0.9) and recorded the aggregation results using a timed aggregation strategy. First, we conduct experiments under a scenario where the device data are identically distributed. As illustrated in Fig. \ref{fig:deviceflow_dropout}(a), the test accuracy under conditions with no dropout and different dropout probabilities exhibited negligible differences. This result is reasonable, suggesting that in an device-cloud collaborative scenario with uniform data distribution, device dropout does not significantly affect the global aggregation computation results. Next, we perform experiments in a scenario where the device data are differentially distributed, with 70$\%$ of the devices having a high proportion of positive samples and 30$\%$ having a high proportion of negative samples. The results under varying dropout probabilities are presented in Fig. \ref{fig:deviceflow_dropout}(b). As the dropout probability increases, the global model's convergence process becomes increasingly unstable, and the test accuracy during the convergence phase also correspondingly decreases. These findings indicate that in device-cloud scenarios with differentially distributed data, simulating device dropout is crucial for assessing the effectiveness of device-cloud algorithms. Consequently, integrating dropout simulation into the simulation platform is essential to accurately mirror real-world application scenarios and to effectively evaluate device-cloud algorithms.

\section{Conclusion}\label{sec:conclusion}
{\color{black}
In this paper, we propose a high-fidelity device simulation platform for real-world device-cloud collaborative computing. The SimDC platform adopts hybrid heterogeneous resources, enabling developers to cost-effectively build a large number of devices for functional testing and obtain physical performance metrics. With the designed hybrid allocation optimization, our platform can minimize the time cost for given simulation requirements. Besides, we design a device behavior traffic controller to simulate real heterogeneous device behaviors. Extensive experiments indicate the effectiveness of our platform and its wide application potential.
}
\section*{Acknowledgments}\label{sec:acknowledgments}

We gratefully acknowledge the strong support and insightful guidance from our supervisor, Jinsong Liao. We thank the OPPO team members for their dedicated contributions: Software Engineers Benchi Xie, Yunku Chen, Hui Wang, Xiangmou Qu, Hongtao Pang, Jin Rui, Chenglong Zhao, Chunying Zhang, Fengli Fan; and Testing Engineers Cheng Jin, Yang Feng, Ruxin Wang, Jingjing Dai, and Ji Chen. We express gratitude to the anonymous reviewers of the ICDCS conference for their constructive comments on this work.

\balance
\bibliographystyle{IEEEtran}
\bibliography{IEEEabrv,references}

\end{document}